\documentclass[12pt]{article}
\usepackage{epsfig}

\newcommand\pubnumber{CERN-TH/2001-007}
\newcommand\pubdate{\today}
\newcommand\hepnumber{hep-ph/0101206}

\def\csumb{CERN, Theory Division\\
CH-1211 Geneva 23, Switzerland}
\def\support{\footnote{Work supported by the European 
Commission TMR programme under the grant ERBFMBICT 983539.}}

\def\Title#1{\begin{center} {\Large\bf #1 } \end{center}}
\def\Author#1{\begin{center}{ \sc #1} \end{center}}
\def\Address#1{\begin{center}{ \it #1} \end{center}}

\newcommand\pubblock{\rightline{\begin{tabular}{l} \pubnumber\\
         \pubdate\\ \hepnumber \end{tabular}}}
\newenvironment{Abstract}{\begin{quotation}  }{\end{quotation}}
\newenvironment{Presented}{\begin{quotation} \begin{center} 
             Presented at the\end{center}
      \begin{center}\begin{large}}{\end{large}\end{center} \end{quotation}}
\def\Acknowledgments{\bigskip  \bigskip \begin{center}
          \large\bf Acknowledgments\end{center}}

\makeatletter
\def\section{\@startsection{section}{0}{\z@}{5.5ex plus .5ex minus
 1.5ex}{2.3ex plus .2ex}{\large\bf}}
\def\subsection{\@startsection{subsection}{1}{\z@}{3.5ex plus .5ex minus
 1.5ex}{1.3ex plus .2ex}{\normalsize\bf}}
\def\subsubsection{\@startsection{subsubsection}{2}{\z@}{-3.5ex plus
-1ex minus  -.2ex}{2.3ex plus .2ex}{\normalsize\sl}}

\renewcommand{\@makecaption}[2]{%
   \vskip 10pt
   \setbox\@tempboxa\hbox{\small #1: #2}
   \ifdim \wd\@tempboxa >\hsize     % IF longer than one line:
       \small #1: #2\par          %   THEN set as ordinary paragraph.
     \else                        %   ELSE  center.
       \hbox to\hsize{\hfil\box\@tempboxa\hfil}
   \fi}

 \def\citenum#1{{\def\@cite##1##2{##1}\cite{#1}}}
\newcount\@tempcntc
\def\@citex[#1]#2{\if@filesw\immediate\write\@auxout{\string\citation{#2}}\fi
  \@tempcnta\z@\@tempcntb\m@ne\def\@citea{}\@cite{\@for\@citeb:=#2\do
    {\@ifundefined
       {b@\@citeb}{\@citeo\@tempcntb\m@ne\@citea\def\@citea{,}{\bf ?}\@warning
       {Citation `\@citeb' on page \thepage \space undefined}}%
    {\setbox\z@\hbox{\global\@tempcntc0\csname b@\@citeb\endcsname\relax}%
     \ifnum\@tempcntc=\z@ \@citeo\@tempcntb\m@ne
       \@citea\def\@citea{,}\hbox{\csname b@\@citeb\endcsname}%
     \else
      \advance\@tempcntb\@ne
      \ifnum\@tempcntb=\@tempcntc
      \else\advance\@tempcntb\m@ne\@citeo
      \@tempcnta\@tempcntc\@tempcntb\@tempcntc\fi\fi}}\@citeo}{#1}}
\def\@citeo{\ifnum\@tempcnta>\@tempcntb\else\@citea\def\@citea{,}%
  \ifnum\@tempcnta=\@tempcntb\the\@tempcnta\else
  {\advance\@tempcnta\@ne\ifnum\@tempcnta=\@tempcntb \else\def\@citea{--}\fi
    \advance\@tempcnta\m@ne\the\@tempcnta\@citea\the\@tempcntb}\fi\fi}
\makeatother

\def\beq{\begin{equation}}
\def\eeq#1{\label{#1}\end{equation}}
\def\eeqn{\end{equation}}

\newenvironment{Eqnarray}%
   {\arraycolsep 0.14em\begin{eqnarray}}{\end{eqnarray}}
\def\beqa{\begin{Eqnarray}}
\def\eeqa#1{\label{#1}\end{Eqnarray}}
\def\eeqan{\end{Eqnarray}}

\let\bar=\overbar

\def\etal{{\it et al.}}

\def\Dslash{\not{\hbox{\kern-4pt $D$}}}
\def\dslash{\not{\hbox{\kern-2pt $\del$}}}

\def\mt{m_t}

\def\msb{{\bar{\ssstyle M \kern -1pt S}}}

\def\lsim{\mathrel{\raise.3ex\hbox{$<$\kern-.75em\lower1ex\hbox{$\sim$}}}}
\def\gsim{\mathrel{\raise.3ex\hbox{$>$\kern-.75em\lower1ex\hbox{$\sim$}}}}

\newcommand{\aS}[1][]{\ensuremath{\alpha_{\mathrm s}^{#1}}}
\newcommand{\tb}[1][]{\ensuremath{\tan^{#1}\!\beta}}

\newcommand{\mW}[1][]{\ensuremath{m_W^{#1}}}
\newcommand{\mb}[1][]{\ensuremath{m_b^{\mathrm{#1}}}}
\newcommand{\omb}{\ensuremath{\overline{m}_b}}
\newcommand{\mtau}{\ensuremath{m_\tau}}
\newcommand{\Dmb}[1][]{\ensuremath{\Delta\mb[#1]}}
\newcommand{\MSUSY}{\ensuremath{M_\mathrm{SUSY}}}
\newcommand{\tbH}{\ensuremath{t\bar{b}H^-}}
\newcommand{\CtbH}{\ensuremath{\bar{t}bH^+}}
\newcommand{\pptbH}{\ensuremath{p\bar{p},\,pp\to\tbH +X}}
\newcommand{\pT}{\ensuremath{p_T}}
\newcommand{\bsg}{\ensuremath{b \to s \gamma}}
\newcommand{\BRbsg}{\ensuremath{{\cal BR}(\bsg)}}
\newcommand{\Hpm}{\ensuremath{H^\pm}}
\newcommand{\mH}{\ensuremath{m_{H^+}}}
\newcommand{\mA}{\ensuremath{m_{A}}}
\newcommand{\hb}{\ensuremath{h_b}}
\newcommand{\hT}{\ensuremath{h_t}}
\newcommand{\mg}{\ensuremath{m_{\tilde{g}}}}

\newcommand{\At}{\ensuremath{A_t}}
\newcommand{\MS}{\ensuremath{\overline{\mathrm{MS}}}}

\newcommand{\eq}[1]{(\ref{#1})} 
\newcommand{\fig}[1]{fig.~\ref{#1}}
\begin{document}
\begin{titlepage}
\pubblock

\vfill
\def\thefootnote{\fnsymbol{footnote}}
\Title{Quantum corrections for the MSSM Higgs
\\[5pt] couplings to SM fermions}
\vfill
\Author{David Garcia\support}
\Address{\csumb}
\vfill
\begin{Abstract}
  Higgs Yukawa couplings to down-type fermions receive, in the MSSM,
  supersymmetric quantum corrections that can be of order 1 for large
  values of \tb, provided $|\mu|\sim\MSUSY$. Therefore, a sensitive
  prediction for observables driven by any of these couplings can only
  be obtained after an all-order resummation of the large corrections.
  We perform this necessary step and show, as an example, the effect
  of the resummation on the computation of the \pptbH\ cross-section
  and on the branching ratio \BRbsg\ at the next-to-leading order.
\end{Abstract}
\vfill
\begin{Presented}
5th International Symposium on Radiative Corrections \\ 
(RADCOR--2000) \\[4pt]
Carmel CA, USA, 11--15 September, 2000
\end{Presented}
\vfill
\end{titlepage}
\def\thefootnote{\arabic{footnote}}
\setcounter{footnote}{0}

\section{Introduction}

The full experimental confirmation of the Standard Model (SM) still
requires the finding of the Higgs boson. The last LEP results, suggesting
a light Higgs of about 115~GeV \cite{LEP115}, are encouraging, but we
will have to wait for the upgraded Tevatron or the LHC to see this
result either confirmed or dismissed.  In any case, there is room for
a extended Higgs sector of various kinds (extra doublets, singlets, even
triplets).

The Higgs sector of the Minimal Supersymmetric Standard Model (MSSM),
well-known nowadays (see \cite{HHG}, for instance), still deserves
further studies. In particular, an interesting topic is the effect of
supersymmetric corrections on the Yukawa interaction, because many
production and decay channels are mediated by these couplings, some of
them at the one-loop level (such as $H\to \gamma\gamma$ or $gg\to H$).

For large \tb\ values, one expects deviations of order 1 of the Yukawa
couplings to down-type fermions from their tree-level values, due to
gluino (SUSY-QCD) and, to a lesser extent, higgsino (SUSY-EW)
radiative effects. This can be seen, for instance, in the computation
of the one-loop correction to the $t\to bH^+$ partial decay rate
\cite{SUSYtbH}, which grows with \tb\ as
$(\alpha_{\mathrm{s,w}}/4\pi)\tb$.  With contributions of order
$(\aS/4\pi)^n\tb[n]$ arising at higher orders in perturbation theory,
the one-loop result can only be meaningful for small \tb\ values.
Let us recall that large \tb\ scenarios, such as those derived from
supersymmetric SO(10) models with unification of the top and bottom
Yukawa couplings at high energies \cite{gut,Dmb}, have become more
appealing since LEP searches for a light neutral Higgs boson, $h$,
started to exclude the low-\tb\ region of the MSSM parameter space.
The latest analyses rule out the MSSM for \tb\ in the range
$0.52<\tb<2.25$, even with maximal stop mixing \cite{LEPtb}.

In this talk we present the resummation of such corrections into the
definition of the bottom Yukawa as a function of the bottom mass,
restoring the reliability of the perturbative series for large \tb.
This ``improved'' formula for the Yukawa is then used in the
evaluation of the \pptbH\ cross-section and of the branching ratio for
\bsg\ at the next-to-leading order (NLO), comparing the result with
the case in which no resummation is made.

\section{Resummation of SUSY corrections}
\label{sect:sum}

Let us briefly explain how such large corrections could arise. The
starting point is the MSSM superpotential. Supersymmetry constrains it
to be holomorphic in the chiral superfields, implying that the
left-handed components of down-type quarks and leptons only couple to
the $H_1$ Higgs doublet, while the left-handed up-type quarks and
leptons only couple to $H_2$. For the third-generation quarks, one has
\begin{equation}
  {\cal L} = - \hb\,\bar{b}b\,H_1^0
             - \hT\,\bar{t}t\,H_2^0 + \cdots
\end{equation}
Soft-SUSY-breaking operators induce the forbidden couplings,
$\bar{b}bH_2^0$ and $\bar{t}tH_1^0$, radiatively. After integrating
out all R-odd particles in the MSSM, one obtains an effective
two-Higgs-doublet model (2HDM) lagrangian
\begin{equation}
  {\cal L}_{\mathrm{eff}}= 
            -(h_b+\Delta h_b^1)\,\bar{b}b H_1^0 
            -(0  +\Delta h_b^2)\,\bar{b}b H_2^0 + \cdots
\end{equation}
As we have argued above, there is a clear motivation for studying the
large \tb\ regime of the MSSM. If \tb\ is large, and after electroweak
symmetry breaking, the $\Delta\hb^2$ term can induce corrections of
order 1 to down-type fermion masses:
\begin{equation}\mb=\hb\,v_1\,\left(
   1+\Delta\hb^1/\hb+\Delta \hb^2/\hb\,\tb\right)\,,
\end{equation}
or conversely, one can express the renormalized bottom Yukawa coupling
as a function of the bottom mass through
\begin{equation}
  \label{eq:hb}
  \hb\,v_1=\frac{m_b}{1+\Delta\hb^1/\hb+\Delta\hb^2/\hb\,\tb}
                      \sim\frac{\mb}{1+\Dmb}\,.
\end{equation}

The set of quantum corrections included in~\eq{eq:hb} are
universal, in the sense that they equally affect all amplitudes
proportional to the bottom Yukawa. To derive~\eq{eq:hb}, one matches
the MSSM to a generic 2HDM at a scale \MSUSY\ of the order of the
relevant soft-SUSY-breaking parameters. Alternatively, within the
MSSM, and using an on-shell renormalization scheme, these corrections
are absorbed into the bottom mass counterterm, $\delta\mb[SUSY]\sim
-\Dmb[SUSY]$.

The quantity \Dmb\ is dominated by SUSY-QCD virtual
effects,\footnote{An analogous quantity, $\Delta\mtau$, can be defined
  for the $\tau$-Yukawa. $\Delta\mtau$, though, receives (generally)
  smaller SUSY-EW contributions.} and at the one-loop level can be
cast into the simple expression \cite{Dmb}
\begin{equation}
\label{eq:Dmb}
  \Dmb\sim\Dmb[\mathrm{SQCD}]=
  \frac{2\aS}{3\pi}\,
  \mu\,\mg\,\tb\,\mathrm{I}(m_{\tilde{b}_1},m_{\tilde{b}_2},m_{\tilde{g}})\,,
\end{equation}
where the function I is the limit of Passarino--Veltman's $C_0$ for
vanishing external momenta.

An interesting property of \Dmb\ is that it does not vanish for
$\MSUSY/\mW\to\infty$. The SUSY-QCD contribution, for instance,
evaluates to $\aS/(3\pi)\,\tb$ in this limit. This should never be
understood as a non-decoupling behaviour of the MSSM, because the
tree-level \hb\ is not an observable. 
If the masses of both the SUSY partners and the non-standard higgses
($H$, $A$, \Hpm) become large, the SUSY radiative corrections to \hb\ 
are cancelled out exactly by one-loop process-dependent corrections.
If \mA\ is not too large with respect to \mW, one can expect large
deviations from the rule
\begin{equation}
  \label{eq:gvm}
  g_{hbb}/g_{h\tau\tau}=\mb/\mtau\,,
\end{equation}
which holds not only in the SM, but also in 2HDM of types I and II
\cite{HHG}.  For the $H$, $A$, neutral higgses, eq.~\eq{eq:gvm} can be
violated even in the decoupling limit, their masses being of the order
of \mA.  This feature could help in distinguishing the MSSM Higgs
sector from a generic type~II 2HDM, specially if correlations among
various Higgs couplings were checked.

\begin{figure}[t]
\begin{center}
\epsfig{file=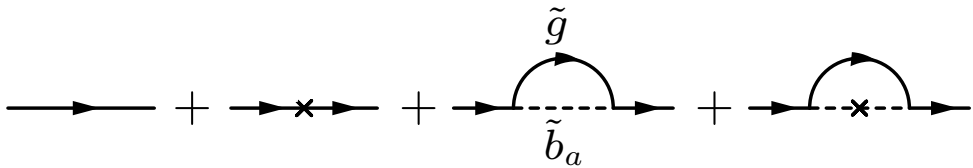,width=0.85\textwidth}
\caption[0]{\label{fig:self} 
  Complete set of Feynman diagrams contributing at order \aS[n]\tb[n]
  to the inverse bottom propagator, in SUSY-QCD with no virtual
  gluons. Dashed and solid internal lines represent sbottom quarks and
  gluinos, respectively.
  A cross denotes, in the second diagram, the insertion of a bottom
  mass-counterterm, and in the last diagram, the counterterm for the
  $\tilde{b}_L\tilde{b}_RH_2^0$ coupling.}
\end{center}
\end{figure}

Remarkably enough, it can be shown that, in mass-independent
renormalization schemes such as the \MS, the whole set of SUSY-QCD
corrections of the form\footnote{The only exception being, for $n=1$,
  the process-dependent one-loop effects that restore the SM
  low-energy limit of the theory for $\mA/\mW\to\infty$.}
$\aS[n]\tb[n]$ are resummed into the above definition for \hb\ 
\cite{cgnw} in eq.~\eq{eq:hb}. The proof involves the consideration of
the perturbative series for the inverse bottom propagator, which can
be used to determine the functional relation between \hb\ and the
bottom pole mass. In a first step one restricts the analysis to the
set of diagrams with no gluons: only those in fig.~\ref{fig:self}
contribute at order \aS[n]\tb[n], higher-loop diagrams being
suppressed either by inverse powers of \tb\ or by $\mb/\MSUSY$
factors. Requiring the inverse propagator to vanish on-shell, one
arrives at~\eq{eq:hb}, apart from $1/\tb$ and $\mb/\MSUSY$ suppressed
quantities. The full proof, that is, after allowing for diagrams
containing virtual gluons, is more delicate. It requires, for
instance, a careful analysis of the infrared behaviour of the extra
diagrams, as $1/\mb$ mass singularities would invalidate the counting
of $\mb/\MSUSY$ powers used in the proof.

%%%%
%%%% Should I include decoupling? 
%%%%
% \begin{figure}[t]
% \begin{center}
% \epsfig{file=nusqcd.eps,height=1in}
% \caption[0]{\label{fig:NUSQCD} Tree-level and one-loop SUSY-QCD non-universal
%   correction to the $h b\bar{b}$, $H b\bar{b}$ and $A
%   b\gamma_5\bar{b}$ effective couplings.}
% \end{center}
% \end{figure}

% In the above-mentioned effective lagrangian picture of the MSSM, and
% after adding the process-dependent SUSY-QCD vertex corrections,
% fig.~\ref{fig:NUSQCD}, the renormalized amplitudes for $h\to
% b\bar{b}$, $H\to b\bar{b}$ and $A\to b\gamma_5\bar{b}$ read:
% %
% \begin{eqnarray}
% \label{eq:renamp}
% \bar{b}b\,h
%   & & \frac{\mb}{v}\,\frac{\sin\aeff}{\cbt}   \frac{1}{1+\Dmb}
%       \left(1-\frac{\Dmb}{\tan\aeff\tb}\right) \nonumber\\[1.5ex]
% \bar{b}b\,H 
%   & &-\frac{\mb}{v}\,\frac{\cos\aeff}{\cbt}   \frac{1}{1+\Dmb}
%     \left(1+\Dmb\,\frac{\tan\aeff}{\tb}\right) \nonumber\\[1.5ex]
% i\,\bar{b}\gamma_5 b A 
%   & & \frac{\mb}{v}\,{\tb} \frac{1}{1+\Dmb}
% \end{eqnarray}
% %
% From eq.~\eq{eq:renamp} one can check that one recovers the SM
% coupling for $h b\bar{b}$ (and $G^0 b\bar{b}$) as $\mA/\mW\to\infty$
% (because $\tan\aeff\to -1/\tb$). 

\section{Prospects for \Hpm\ searches at hadron colliders}

As a first example of the use of eq.~\eq{eq:hb}, we are going to
consider the MSSM associated production of a charged Higgs boson,
\Hpm, with top and bottom at hadron colliders, presenting results for
both the Tevatron and the LHC \cite{bggs}. From our point of view, the
relevance of this channel is due to its ability to test the charged
Higgs coupling to the third-generation quarks and leptons. Any
information obtained about these couplings could provide valuable
hints on the exact nature of the Higgs sector.

After the LEP shutdown,
% direct LEP limit, $\mH>79.9$~GeV, for a generic 2HDM \Hpm\ 
% decaying exclusively into $\tau\nu_\tau$ \cite{LEPHpm}.
charged Higgs searches concentrate on the Tevatron results.  Both
direct and indirect Tevatron analyses have been limited to the region
$\mH<\mt-\mb$, placing constraints on the \mH--${\cal BR}_{t\to b
  H^+}$ plane \cite{TeVHpm}, which are usually translated to the
\mH--\tb\ plane once the relevant MSSM parameters are fixed
\cite{tbmh}.
%
% Finally, the bound coming from the rare process \bsg, which for a pure
% type~II 2HDM places a tight constraint in the \mH--\tb\ plane,
% although in the MSSM lighter \Hpm\ masses can be traded for a
% constraint in the \At--$\mu$ set \cite{}.

% Indirect searches in B-physics observables (apart from \bsg) are also
% promising, specially in $b \to s\,l^+l^-$ and $b\to c\,\tau\,\nu_\tau$
% \cite{BH}.

Beyond the kinematical limit for $t\to b H^+$, apart from the one
considered in this talk, there are two other promising production
channels: pair production \cite{ppHH} and associated production with a
$W$ boson \cite{ppHW}.  
The work presented here on \pptbH\ adds, with respect to previous
analyses \cite{Gunion,Borzumati,Coarasa,Kfactor,morepptbH}, a
resummation of the leading ---in powers of \aS\tb, for $\tb\gsim
10$--- SUSY radiative corrections (both strong and electroweak) and a
estimation of the off-shell effects. It will be presented in full
detail in \cite{bggs}, including a complete signal and background
analysis.

\subsection{Cross-section computation and results}

\begin{figure}[t]
\begin{center}
  \epsfig{file=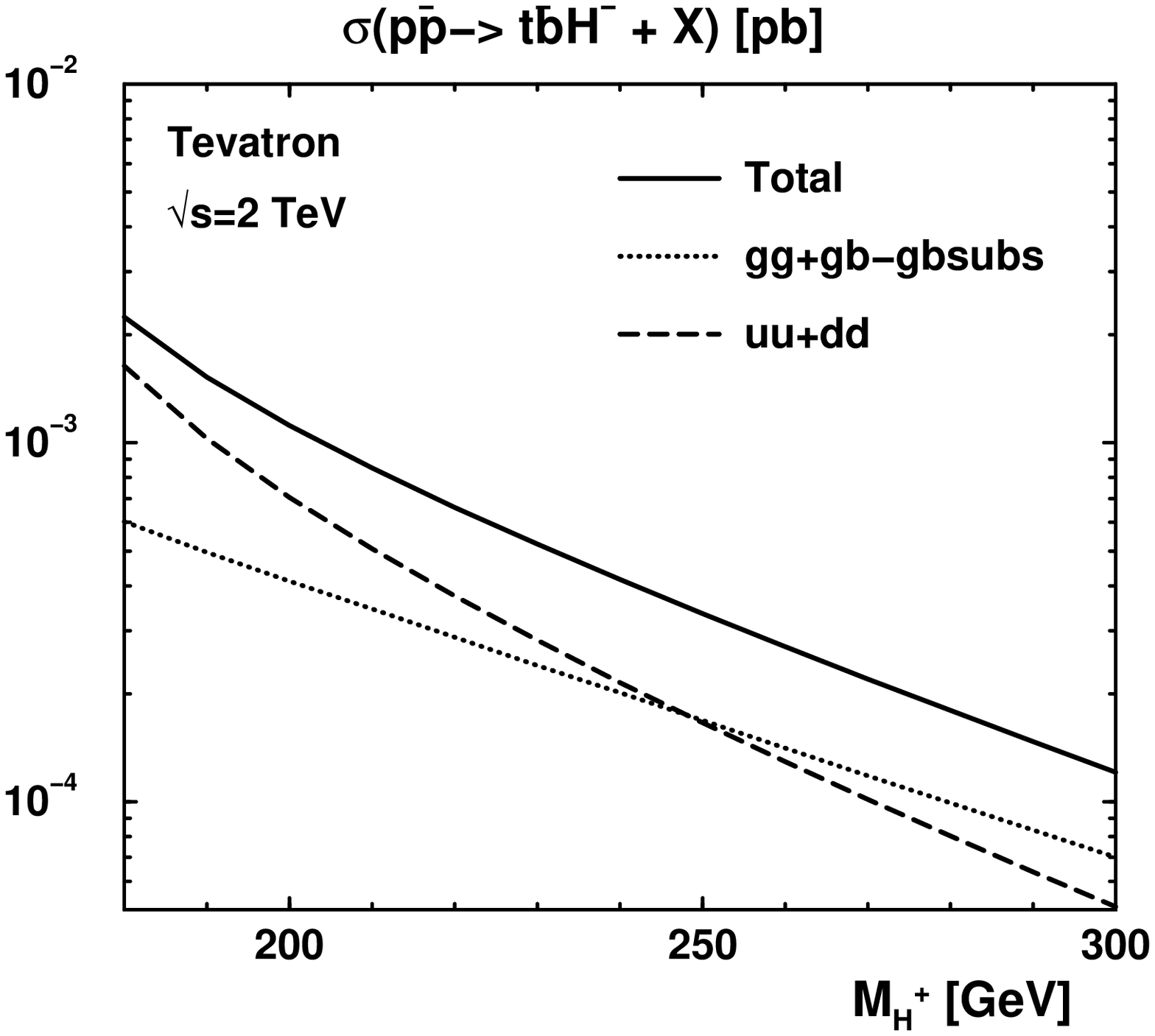,width=0.43\textwidth}\hspace{0.8cm}
  \epsfig{file=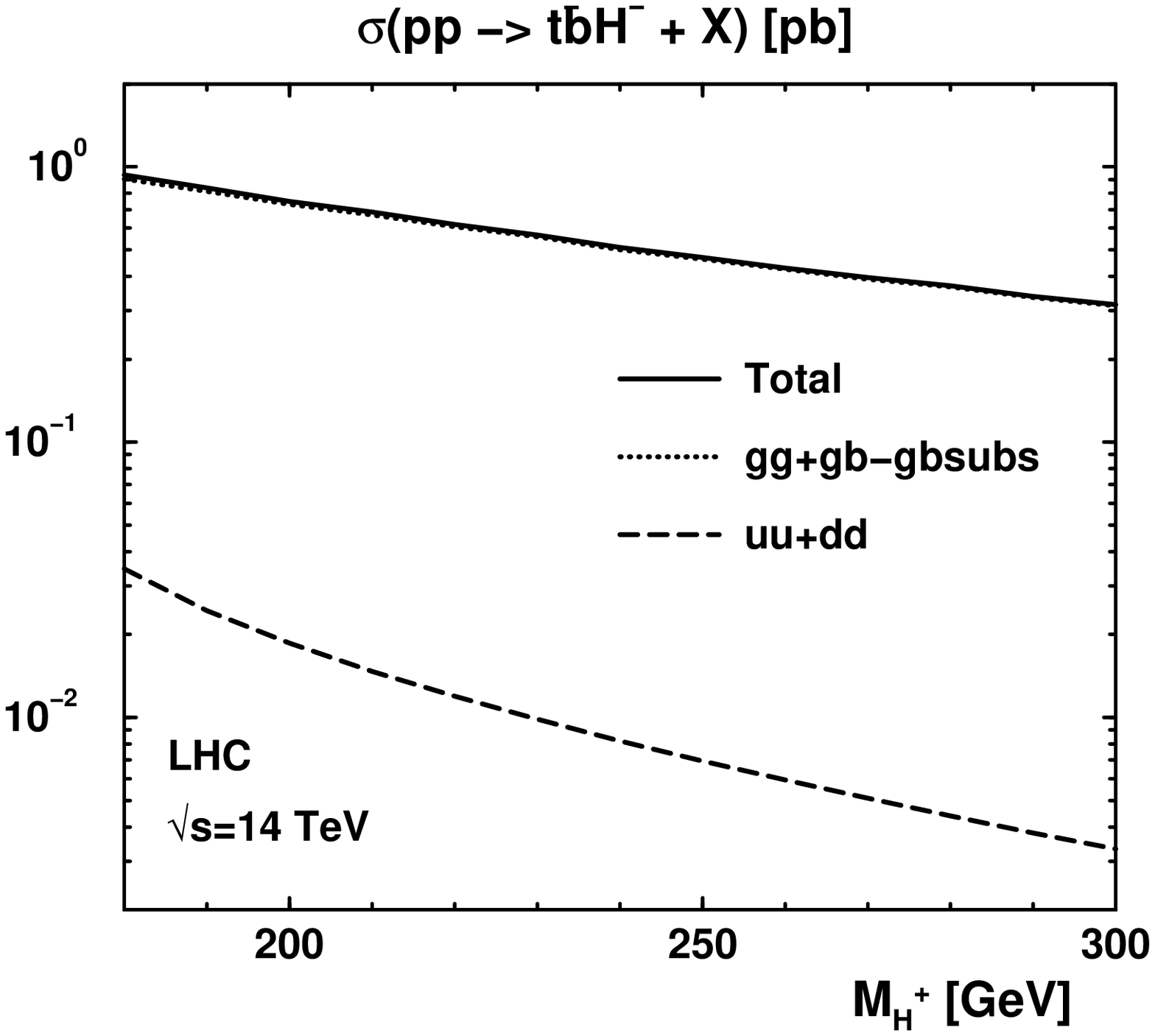,width=0.43\textwidth}
\caption[0]{\label{fig:reach}
  The \pptbH\ cross-section at the Tevatron Run~II (left) and at the
  LHC (right), for $\mu=-200$~GeV, $\tb=30$ and
  $\mg=m_{\tilde{t}_1}=m_{\tilde{b}_1}=A_b=\At=500$~GeV. The dashed
  curve corresponds to the $q\bar{q}$ annihilation channel, the dotted
  curve to the sum of the $gg$-initiated and $gb$-initiated channels,
  after subtracting double counting in the $gb$ channel. The solid
  curve is the sum of all channels, $q\bar{q}$, $gg$ and $gb$.}
\end{center}
\end{figure}

At the parton level, the reactions \pptbH\ proceed through three main
channels\footnote{We shall omit the charge-conjugate process,
  $p\bar{p},\,pp\to\CtbH +X$, for the sake of brevity. Including this
  process just amounts to multiplying our cross-section by a factor of
  2.}: i) $q\bar{q} \to \tbH$, with $q=u,d$ (the $s$ contribution can
be safely neglected), a channel only relevant to the Tevatron
\cite{Borzumati,Coarasa}; ii) $gg \to \tbH$, dominant at the LHC, and
also at the Tevatron for increasing \Hpm\ masses
\cite{Borzumati,Coarasa}.
Since the bottom mass, \mb, is small with respect to the energy of the
process, parton distribution functions (PDFs) for $b$-quarks have to
be introduced, allowing for the resummation of collinear logs
\cite{Olness}. This provides an extra $bg \to tH^-$ channel
contributing to the cross-section.  Contrary to i) and ii), in this
case the final state contains at most three high-\pT\ $b$-quarks and,
therefore, $bg$-initiated processes cannot be detected by using four
high-\pT\ $b$-tagging.  Once a PDF for the $b$-quarks is used, there
is some amount of overlap between $bg$- and $gg$-initiated amplitudes,
which has to be removed \cite{Olness,Dicus}. To this end, we follow
here the method described in ref.~\cite{Dicus}, straight\-forwardly
translated to the \tbH\ final state case (see also \cite{Borzumati}).
Figure~\ref{fig:reach} shows the relative relevance of the various
channels to both the Tevatron Run~II and the LHC, as explained above.
The solid curve can be used to get a rough estimate of the reach of
the process \pptbH\ in the search for a MSSM charged Higgs boson in
these machines for the given parameters (see the caption).

\begin{figure}[t]
\begin{center}
\epsfig{file=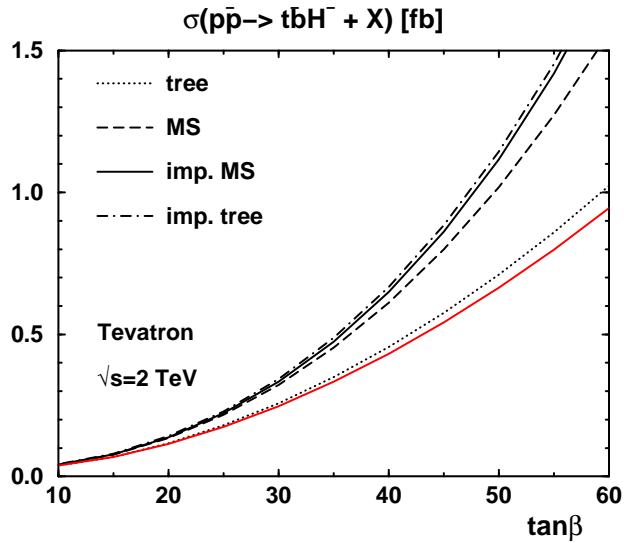,height=0.35\textheight}
\caption[0]{\label{fig:impB} 
  Various approximations to $\sigma(p\bar{p}\to \tbH + X)$ at the
  Tevatron Run~II, as a function of \tb, for a charged Higgs mass of
  250~GeV. The remaining MSSM parameters are set as in
  \protect{\fig{fig:reach}}. Shown are the tree-level (dotted line),
  the one-loop \MS\ (dashed line), the improved or resummed \MS\ (top
  solid line and bottom red solid line, the latter for $\mu=200$~GeV
  and $\At=-500$~GeV) and the improved tree-level (dot-dashed line)
  results.}
\end{center}
\end{figure}

The amplitude for \pptbH\ is, for large \tb, approximately
proportional to the Yukawa coupling of the bottom quark and, as
explained above, receives supersymmetric quantum corrections that can
be of order 1. An analysis of the reach of \pptbH\ in \Hpm\ searches
demands the appropriate inclusion and resummation (using \eq{eq:hb})
of such corrections in the computation of the cross-section.
For the present work we have also included the full off-shell SUSY-QCD
and SUSY-EW corrections to the $H^+\bar{t}b$ vertex and to the fermion
propagators, although \Dmb\ in \eq{eq:hb} is the only correction
contributing at order $(\aS/4\pi)^n\tb[n]$ and thus dominates for large
\tb.
In fact, the approximation of neglecting vertex and propagator
corrections in the cross-section, which we call ``improved
Born'' approximation, is really justified in that region (see
\fig{fig:impB}).

We disregard virtual supersymmetric effects on the $gqq$ and $ggg$
vertices and on the gluon propagators. We expect those to be of order
$(\aS/4\pi)\cdot (\sqrt{s}/\MSUSY)$, with no \tb\ enhancement, and
thus suppressed both by a loop factor (any reasonable choice for
$\aS(Q)$ will be small) and by a MSSM form factor coming from the loop
integrals. Therefore, we can neglect these contributions as we are
only considering large \tb\ values. Besides, the cross-section for the
signal is much smaller for \tb\ close to 1, so our approximation is
well justified.

The only other source of potentially large radiative corrections is,
of course, standard QCD. At least one group is currently addressing
the NLO QCD correction to $p\bar{p},\,pp \to bbH + X$, which can
provide a good guess for the sign and size of the corrections in
\pptbH. In the meantime, we can parametrize our ignorance by using a
K-factor ranging between 1.2 and 1.5 \cite{Dawson,Kfactor,Spira}. Once
the exact value of K will be known, it will be easy to conveniently
rescale our plots to take into account the effect of the gluon loops.
The only QCD corrections we do incorporate to the cross-section are
those related to the running of $\aS(Q)$ and $\omb(Q)$.\footnote{In
  the $t\to bH^+$ decay rate, this actually accounts for most of the
  QCD virtual effects (for $Q=\mt$) \cite{QCDtbH}.}  We choose to work
with equal renormalization, $\mu_R$, and factorization, $Q$, scales
fixed at $\mu_R=Q=\mt+\mH$.

Concerning the method employed to compute the squared matrix elements,
we have made intensive use of the package CompHEP \cite{CompHEP}, for
both the signal and background processes. Although CompHEP is only
able to deal with tree-level calculations, we have managed to add the
supersymmetric corrections to the \tbH\ vertex and to the fermion
propagators in the following way: first, we have modified CompHEP's
Feynman rules to allow for the most general off-shell \tbH\ vertex,
then we have let CompHEP reckon the squared matrix elements and dump
the result into REDUCE
%\cite{} 
code. At this point, we have inserted expressions for the coefficients
of the off-shell \tbH\ vertex that include the one-loop off-shell
supersymmetric corrections to the vertex itself and to the off-shell
fermion propagators and fermionic external lines.\footnote{We shall
  not write down here the analytic expressions for the renormalized
  vertex and propagators. They can easily be derived by just
  generalizing previous on-shell calculations, such as those for $t\to
  bH^+$ \cite{SUSYtbH}.}  Only half the renormalization of an internal
fermion line has to be included, the other half being associated to
the $gqq$ vertex. This procedure has allowed us to estimate the
relative size of the off-shell effects in the signal cross-section,
which never exceeds the few per cent level.

In \fig{fig:impB}, we compare the above various approximations to the
$p\bar{p}\to \tbH + X$ cross-section at the Tevatron Run~II. The
curves correspond to the total cross-section, as a function of \tb,
for a centre-of-mass energy of 2~TeV and a charged Higgs mass of
250~GeV.
The tree-level result is given by the dotted line; it grows almost
quadratically with \tb. After including the \MS\ off-shell one-loop
supersymmetric corrections in the \tbH\ vertex, in the internal
fermion propagators and in the external fermion lines, one obtains the
dashed line. For the chosen parameters, that is $\mu=-200$~GeV,
$\mg=m_{\tilde{t}_1}=m_{\tilde{b}_1}=A_b=\At=500$~GeV, the overall
correction turns out to be positive, as it is driven by the SUSY-QCD
correction in \eq{eq:hb}, which is positive for negative $\mu$.
The resummation of the order \aS[n]\tb[n] supersymmetric corrections
further increases the result up to the top solid curve, labelled
``improved'' \MS. The effect is not dramatic because $\mu$ is sizeably
smaller than the rest of the relevant soft-SUSY-breaking parameters,
namely the gluino and sbottom masses. To illustrate the possibility of
a suppression of the cross-section due to virtual supersymmetric
effects, we also plot the resummed result for the same parameters but
taking $\mu=200$~GeV and $\At=-500$~GeV, which corresponds to the
bottom (red) solid line. It does not differ much from the tree-level
because of a partial cancellation of the correction due to the effect
of the resummed high-order terms in \eq{eq:hb}.
Finally, the dot-dashed curve is obtained by just replacing the
$\omb(Q)$ in the tree-level approximation to the cross-section with
$\omb(Q)/(1+\Dmb)$, as suggested by \eq{eq:hb}. This ``improved''
tree-level constitutes a fairly good approximation to the complete
resummed \MS\ result (top solid curve).
Similar conclusions apply for $pp\to\tbH + X$ at the LHC.
%%Similar conclusions apply for the $pp\to\tbH + X$ cross-section at the
%%LHC.

\section{\bsg\ and supersymmetry with large \tb}

The computation of the \bsg\ branching ratio at the NLO in the MSSM is
clearly a complicated matter \cite{cdgg,NLObsg}, and completely general
expressions have not yet been derived. Nevertheless, it turns out that
the leading \aS[n]\tb[n+1] corrections can be calculated and resummed
to all orders in perturbation theory by replacing the tree-level
Yukawa of the bottom quark by eq.~\eq{eq:hb} in the Wilson
coefficients of the leading-order (LO) computation \cite{bsgtb}.

\begin{figure}[t]
\begin{center}
  \epsfig{file=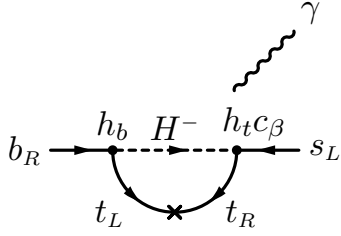,height=1.2in}
\caption[0]{\label{fig:Hbsg} Dominant, in powers of \tb, LO
  contribution to the \bsg\ branching ratio in the 2HDM.}
\end{center}
\end{figure}

To show how this procedure works, let us start by identifying the
dominant one-loop Feynman diagrams. In a type~II 2HDM, the diagram
contributing to the highest \tb\ power is that in fig.~\ref{fig:Hbsg},
with the exchange of a virtual charged Higgs in the loop, which is
proportional to $\hb\cdot\hT\cos\!\beta$ and, therefore, of order
\tb[0]. Substituting a chargino %%, $\chi^-$, 
by the charged Higgs, one obtains the leading diagram in the MSSM,
which is proportional to $\hb\cdot\hT$ and of order \tb. We have
checked, using the formulae in \cite{cdgg}, that the leading ---in
\tb--- NLO corrections follow from the replacement $\hb\to -\hb\Dmb$
(expanding and truncating at first order in \aS) in the LO
contributions associated to the above-mentioned diagrams. These NLO
corrections are to be associated with the insertion of the counterterm
for the bottom Yukawa in the LO, and thus have their origin in a
one-loop diagram. The resulting effect is of order \aS\tb[2] for the
chargino loop, and \aS\tb\ for the $H^-$ loop.

There is one additional source of \tb-enhanced corrections in the
charged Higgs diagrams, which is not related to the bottom mass
counter\-term~\cite{cdgg}: while the tree-level $H^+ \bar{t}_R s_L$
vertex is suppressed by $1/\tb$, this suppression is absent at the
one-loop level, so that the NLO charged-Higgs contribution to \BRbsg\ 
is \tb-enhanced with respect to the LO one. No enhancement occurs in
higher-loop diagrams, which are suppressed either by $1/\tb$ powers or
by $\mb/\MSUSY$ factors \cite{cgnw}.
%%
%% Smallness of the strange quark mass -> need for top Yukawa
%% Comparison with t->bH
%%

Now that all sources of \tb-enhanced terms have been identified, it is
an easy matter to proceed with the improvement of the NLO expressions,
that is, to resum all terms of order \aS[n]\tb[n+1]. Just replace, in
the LO expressions, \hb\ by $\hb/(1+\Dmb)$, and remove double
counting, i.e. $-\hb\Dmb$ terms, in the NLO formulae (see
ref.~\cite{bsgtb} for a detailed description of the procedure). This
is enough to extend the validity of the calculation presented in
ref.~\cite{cdgg} to large values of \tb.

\begin{figure}[t]
\centerline{
\psfig{figure=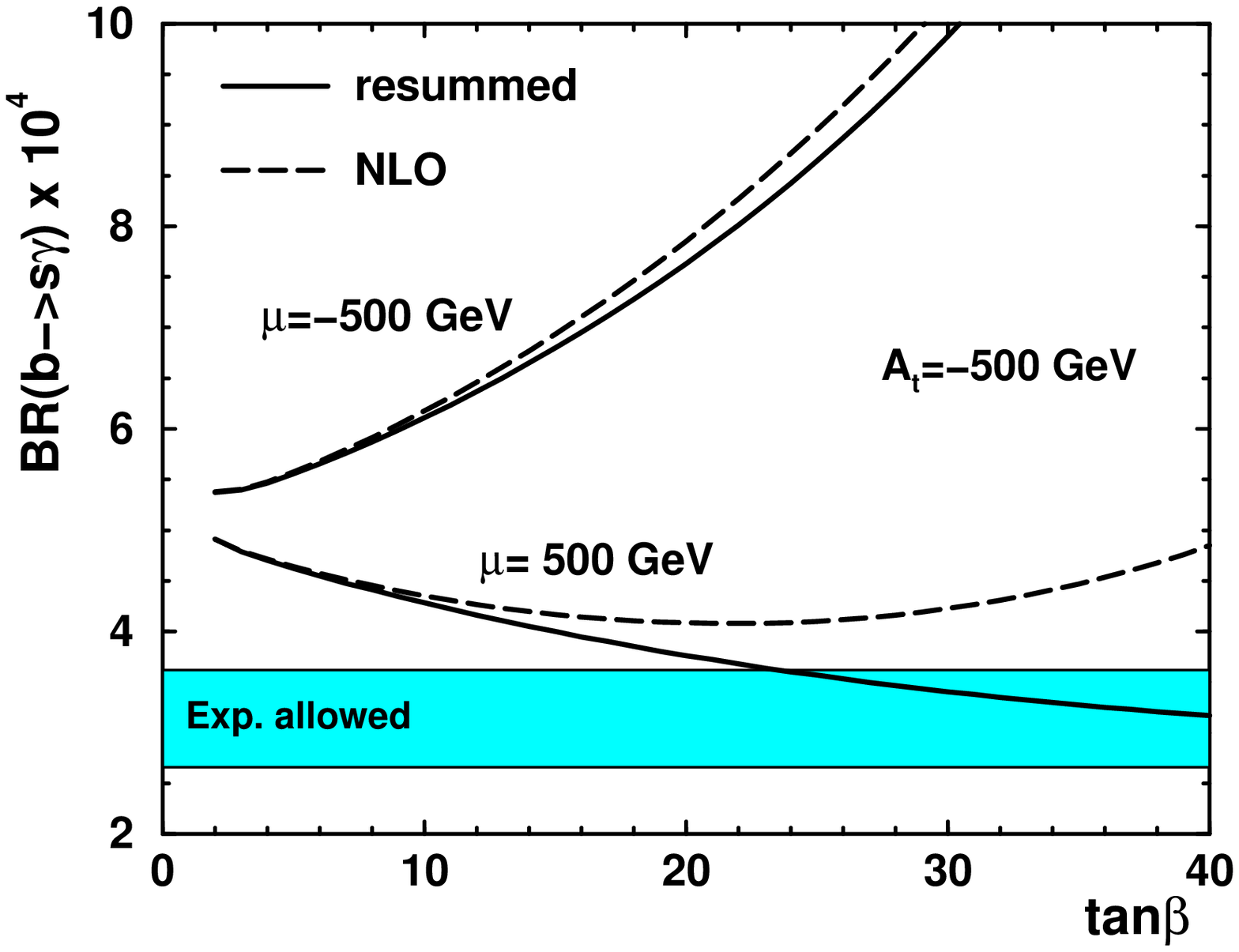,width=8cm}}
\caption{
  Comparison of the NLO theoretical prediction for \BRbsg\ of
  ref.~\cite{cdgg} before (solid line) and after (dashed line)
  performing a resummation of the dominant supersymmetric corrections
  of order \aS[n]\tb[n+1], as explained in the text.
  The charged-Higgs boson mass is $200$~GeV and the light stop mass is
  $250$~GeV. The values of $\mu$ and \At\ are indicated in the plot,
  while $M_2$, the gluino, heavy-stop and down-squark masses are set
  at $800$~GeV.}
\label{fig:bsg}
\end{figure}

The quantitative effect of the improvement can be assessed
from \fig{fig:bsg}, where we compare the NLO theoretical prediction
for \BRbsg\ of ref.~\cite{cdgg} with (solid line) and without (dashed
line) including the all-order resummation of the dominant \tb-enhanced
radiative corrections, for typical values of the supersymmetric
parameters.
We use a negative value of $\At=-500$~GeV at low energies, with
$\mg\At<0$. This choice is inspired in supergravity models, where this
sign relation holds unless the boundary value of \At\ at the
high-energy input scale is one order of magnitude larger than the
gaugino soft-supersymmetry-breaking mass parameters~\cite{gut,iblop}.
For the experimental measurement of the \bsg\ branching ratio, we use
the combined result of CLEO \cite{CLEO} and ALEPH \cite{ALEPH},
$\BRbsg=(3.14\pm 0.48)\times 10^{-4}$.
Owing to cancellations among the various contributions, the relative
size of the effect turns out to be sizeable only for $\mu\At<0$: for
the set of parameters used in \fig{fig:bsg}, it decreases the NLO
result by 20\% at $\tb\simeq 30$.

Notice that, with our sign conventions, positive values of $\mu$ are
necessary in order to obtain correct values for \BRbsg, even after
considering higher-order effects, within minimal supergravity models,
for which, as explained in the above paragraph, the sign of \At\ at
low energies tends to be negative.
% For $\At<0$, a negative sign for the product $\mu\At$ is
% preferred at the NLO, as it is also the case at the LO. 
This is in contradiction with the results of ref.~\cite{db}. We
believe sign errors in the charged Goldstone and Higgs couplings to
stop and down-like squarks in the published version of \cite{cdgg} are
at the origin of this discrepancy (see \cite{bsgtb}).\footnote{The
  authors of \cite{cdgg} have independently detected these sign
  errors, and posted a corrected version of the paper to the hep-ph
  archive.}

\section{Conclusions}

Motivated by the latest LEP analyses ruling out \tb\ values around 1
\cite{LEPtb} and by large \tb\ supersymmetric SO(10) models, we have
analysed the Yukawa couplings of down-type fermions in the MSSM, which
receive potentially large supersymmetric corrections when \tb\ is
large.
We claim that, in observables where these couplings are relevant, a
resummation of the leading ---in powers of \tb--- subset of
corrections is needed if the observables are to be computed for large
\tb. We perform such resummation, which is essentially independent of
the particular process under consideration, and explore its numerical
impact on two exemplary processes: the \pptbH\ cross-section and the
branching ratio of \bsg.

\Acknowledgments 

The author thanks C.E.M~Wagner, M.~Carena, J.~Sol{\`a}, M.J.~Herrero
and A.~Pineda for useful discussions, U.~Nierste and J.~Guasch for
revising the text, J.~Guasch and A.~Belyaev for their collaboration in
the computation of the \pptbH\ cross-section.

{\bigskip  \bigskip \begin{center}
          \large\bf Note added\end{center}}
After presentation of this talk, and concerning the \bsg\ branching
ratio, similar formulae for the improvement of the NLO computation in
the MSSM with large \tb\ have been given in \cite{dgg}.

\end{document}